\documentclass[a4paper,11pt]{article}

\pdfoutput=1
\usepackage{jcappub}

\usepackage{bbm}
\usepackage{mathrsfs}
\usepackage{slashed}
\usepackage{caption}
\usepackage{epstopdf}
\usepackage[normalem]{ulem}
\usepackage[bottom]{footmisc}
\usepackage{subcaption}
\usepackage{bbold}
\usepackage{titlesec}
\usepackage{threeparttable}
\usepackage{booktabs}
\usepackage{changepage}
\usepackage[utf8]{inputenc}
\usepackage{dsfont} 
\usepackage{grffile}
\usepackage{graphicx}
\usepackage{dcolumn}
\usepackage{bm}
\usepackage{amssymb}
\usepackage{setspace}
\usepackage{amsmath}
\usepackage{array}
\usepackage{float}
\usepackage{lmodern}
\usepackage{multirow}
\usepackage{soul}
\usepackage{braket}
\usepackage{comment}
\usepackage[draft]{pgf}
\usepackage{adjustbox} 
\usepackage{xspace} 
\usepackage{url}
\usepackage{xcolor}
\usepackage{orcidlink}
\usepackage{chngcntr}
\usepackage{indentfirst}

\allowdisplaybreaks

\newcommand{\be}{\begin{equation}}
\newcommand{\ee}{\end{equation}}

\newcommand{\bea}{\begin{eqnarray}}
\newcommand{\eea}{\end{eqnarray}}

\newcommand{\mrm}{\mathrm}

 \def \dd {{\rm d}}
\arxivnumber{DESY-25-188}

\title{Stellar Superradiance and Low-Energy Absorption in Dense Nuclear Media}

\author{
Zhaoyu Bai$^{a}$,
Vitor Cardoso$^{b,c}$,
Yifan Chen$^{d,e}$,
Yuyan Li$^{a}$,
Jamie I.~McDonald$^{f}$,
and Hyeonseok Seong$^{g,h}$
}

\affiliation{
$^a$Department of Particle Physics and Astrophysics, Weizmann Institute of Science, 234 Herzl St, Rehovot 7610001, Israel\\
$^b$Center of Gravity, Niels Bohr Institute, Blegdamsvej 17, 2100 Copenhagen, Denmark\\
$^c$CENTRA, Departamento de F\'{\i}sica, Instituto Superior T\'ecnico (IST), Universidade de Lisboa (UL), Avenida Rovisco Pais 1, 1049 Lisboa, Portugal\\
$^d$State Key Laboratory of Dark Matter Physics, Tsung-Dao Lee Institute, Shanghai Jiao Tong University, Shanghai 200240, China\\
$^e$Key Laboratory for Particle Astrophysics and Cosmology (MOE) \& Shanghai Key Laboratory for Particle Physics and Cosmology, Shanghai Jiao Tong University, Shanghai 200240, China\\
$^f$Department of Physics and Astronomy, University of Manchester, Oxford Road, Manchester M13 9PL, United Kingdom\\
$^g$Department of Physics and Jockey Club Institute for Advanced Study, The Hong Kong University of Science and Technology, Clear Water Bay, Kowloon, Hong Kong S.A.R., P.R.China\\
$^h$Deutsches Elektronen-Synchrotron DESY, Notkestr.\ 85, 22607 Hamburg, Germany
}

\emailAdd{zhaoyu.bai@weizmann.ac.il}
\emailAdd{vitor.cardoso@nbi.ku.dk}
\emailAdd{chen.yifan@sjtu.edu.cn}
\emailAdd{yuyan.li@weizmann.ac.il}
\emailAdd{jamie.mcdonald@manchester.ac.uk}
\emailAdd{hseong@ust.hk}

\abstract{Ultralight bosons such as axions and dark photons are well-motivated hypothetical particles, whose couplings to ordinary matter can be effectively constrained by stellar cooling. Limits on these interactions can be obtained by demanding that their emission from the stellar interior does lead to excessive energy loss. 
An intriguing question is whether the same microphysical couplings can also be probed through neutron star superradiance, in which gravitationally bound bosonic modes grow exponentially by extracting rotational energy from the star. Although both processes originate from boson–matter interactions, they probe very different kinematic regimes. Stellar cooling probes boson emission at thermal wavelengths, while superradiance is governed by modes whose wavelength is comparable or greater than the size of the star.
Previous work has attempted to relate the microphysical neutron–nucleon scattering and inverse-bremsstrahlung absorption rates directly to the macroscopic growth rate of superradiant bound states. In this work, we re-examine this connection and show that a naive extrapolation of the microphysical absorption rate to the superradiant regime would imply superradiant rates comparable to astrophysical timescales characterised by pulsar spindown. These naive rates are especially high for vector fields. However, we demonstrate that this conclusion changes once collective multiple-scattering effects in dense nuclear matter are taken into account. Repeated nucleon collisions modify the effective low-energy absorption experienced by the bosonic bound state, strongly suppressing the rate relevant for superradiance.

\bigskip

}

\begin{document}

\maketitle
\flushbottom

\section{Introduction}

Hypothetical bosons such as axions and dark photons are well-motivated extensions beyond the Standard Model (SM) of particle physics. The QCD axion provides an elegant solution to the strong CP problem~\cite{Peccei:1977hh,Weinberg:1977ma,Wilczek:1977pj,Kim:1979if,Shifman:1979if,Dine:1981rt,Zhitnitsky:1980tq,Preskill:1982cy,Abbott:1982af,Dine:1982ah}, while broader families of axion-like particles and dark photons arise generically in fundamental theories with extra dimensions~\cite{Svrcek:2006yi,Abel:2008ai,Arvanitaki:2009fg,Goodsell:2009xc}. A wide range of astrophysical observations place strong constraints on such bosons. In particular, stellar-cooling (SC) arguments exclude large regions of parameter space for boson masses below the eV scale by requiring that new dissipative channels do not exceed energy loss from  Standard Model emission processes~\cite{Raffelt:1996wa,Raffelt:1999tx,Caputo:2024oqc}. These bounds come from a diverse of environments, including neutron star (NS) and supernovae cooling constraints on couplings to nucleons~\cite{Buschmann:2021juv,Carenza:2019pxu,Lella:2023bfb,Dietrich:2019shr,Harris:2020qim}, as well as white dwarf~\cite{Giannotti:2017hny} and red giant~\cite{Capozzi:2020cbu} limits on interactions with electrons.
Axions emitted from these astrophysical sources can be further converted into gamma-ray photons in astrophysical magnetic fields through axion-photon coupling. This process leads to significant constraints on the axion parameter space using gamma-ray telescopes \cite{Fiorillo:2025gnd,Candon:2025sdm,Fiorillo:2021gsw,Lella:2024hfk,Manzari:2024jns,Lecce:2025dbz}.

Beyond relativistic emission, ultralight bosons can also form gravitational bound states around rapidly spinning compact objects. These ``gravitational atoms" extract angular momentum from the host star or black hole through superradiance~\cite{Penrose:1971uk,ZS,Detweiler:1980uk,Cardoso:2005vk,Dolan:2007mj,Brito:2015oca,Bao:2022hew}, leading to exponential growth of the bosonic cloud. Black hole superradiance has already provided stringent constraints on light bosons whose self-interactions do not quench the instability~\cite{Arvanitaki:2010sy,Yoshino:2012kn,Gruzinov:2016hcq,Fukuda:2019ewf,Baryakhtar:2020gao,Omiya:2020vji,Omiya:2022mwv,Omiya:2022gwu,Bao:2022hew,Cheng:2022jsw,Chen:2023vkq,Guo:2024dqd,Witte:2024drg,Guo:2025dkx,Collaviti:2024mvh,Takahashi:2024fyq,Aurrekoetxea:2024cqd,Miller:2025yyx,Xie:2025npy}. For NSs, millisecond pulsars offer an obvious arena in which to test superradiance: they are rapidly rotating, long-lived, and contain dense nuclear matter that can dissipate bosonic excitations. This has motivated the study of stellar superradiance (SS) as a potential probe of boson couplings through precise pulsar spin measurements~\cite{Cardoso:2015zqa,Cardoso:2017kgn,Day:2019bbh,Kaplan:2019ako,Chadha-Day:2022inf,Spieksma:2025sda,Sirkia:2025xpv}.

Although superradiance and stellar cooling can both originate from boson–nucleon interactions, the relevant boson wavelength scales are completely different. Stellar cooling involves the emission of relativistic bosons from individual nucleon–nucleon collisions. In contrast, superradiance is governed by the long-wavelength gravitational bound state of the boson~\cite{Detweiler:1980uk,Brito:2015oca,Baryakhtar:2017ngi,Chen:2022kzv,Chen:2023vkq}, whose spatial extent is set by the Bohr radius of the gravitational atom and can exceed microphysical nuclear scales by many orders of magnitude. This raises a natural question: how should one relate or compare the constraints from these two processes when they arise from the same underlying microphysical interaction?

In this work, we critically re-examine the connection between the microphysical scattering amplitudes and the effective absorption experienced by superradiant bound states. While both processes originate from similar microphysics, superradiance requires treating the boson as a non-relativistic gravitational bound state. We show that a naive extrapolation of the microphysical absorption rate to the superradiant regime would predict rapid superradiant growth, at a level comparable to that inferred from NS spin-down measurements, particularly for vector fields. However, we demonstrate that this conclusion changes qualitatively and quantitatively once multiple-scattering (collective) medium effects inside dense nuclear matter are taken into account. Repeated nucleon collisions strongly suppress the low-energy absorption of long-wavelength bosonic modes~\cite{Friman:1979ecl,Iwamoto:1984ir,Brinkmann:1988vi,Raffelt:1991pw,Raffelt:1996di,Raffelt:1996wa,Shternin:2013pya,Shternin:2020igy,BaymPethick}, suggesting other long-wavelength channels must be sought to enable efficient stellar superradiance.

\section{Stellar Cooling}
\label{sec:cooling}

Stellar cooling provides a powerful method to constrain feebly interacting particles, including ultralight bosons. Emission of such particles from stellar interiors would drain energy at a rate exceeding SM processes dominated by neutrino losses, as inferred from astrophysical observations and stellar-evolution simulations~\cite{Raffelt:1996wa,Raffelt:1999tx}. NSs offer the strongest constraints on couplings to neutrons due to their extreme densities, $n \sim 10^{38}~\mathrm{cm}^{-3}$, and high core temperatures, $T \sim 10^9~\text{K}$~\cite{Iwamoto:1984ir,Brinkmann:1988vi,Iwamoto:1992jp,Buschmann:2021juv}.

Axions $(a)$ naturally couple to neutrons $(\psi)$ through the derivative interaction
\be
\frac{g_a}{2 m_\psi} \partial_{\mu} a \bar{\psi} \gamma^{\mu} \gamma^{5} \psi,\label{eq:LaN}\ee
where $g_a$ is the coupling strength and $m_\psi$ is the neutron mass. Dark photons $(V_\mu)$ admit similar dimension-5 operators through magnetic dipole moment (MDM) and electric dipole moment (EDM) portals~\cite{Hoffmann:1987et,Dobrescu:2004wz}:
\be\begin{split}
\frac{g_{\mathrm{MDM}}}{4 m_\psi} V_{\mu \nu} \bar{\psi} \sigma^{\mu \nu} \psi,\\
\frac{g_{\mathrm{EDM}}}{4m_\psi} V_{\mu \nu} \bar{\psi}\sigma^{\mu \nu} i \gamma^5 \psi,\label{eq:EMDM}
\end{split}\ee
with coupling strengths $g_{\mathrm{MDM}}$ and $g_{\mathrm{EDM}}$, respectively, and $V_{\mu\nu}$ the dark photon field strength. In the non-relativistic limit, these bosons couple mainly to neutron spin currents: the axion via $\vec{\nabla} a$, and dark photons via $\vec{\nabla}\times\vec{V}$ (MDM) or $\partial_t \vec{V}$ (EDM)~\cite{Graham:2013gfa,Budker:2013hfa,Graham:2017ivz,Jiang:2021dby,Chen:2021bdr,Jiang:2024boi}. Other interactions, such as Yukawa couplings, gauge couplings, or kinetic mixing, are more strongly constrained by fifth-force experiments than by stellar cooling and are not considered here.

\begin{figure}[t]
\centering
    \includegraphics[scale=1]{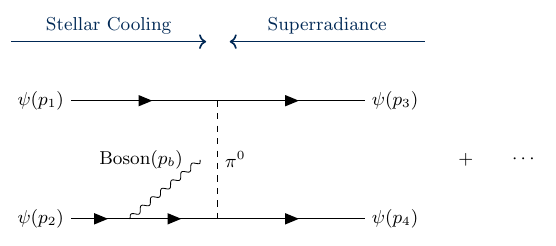}
    \caption{One of the eight Feynman diagrams for nucleon–nucleon bremsstrahlung, involving four external neutrons $(\psi)$ with momenta $p_i$ and one boson with momentum $p_b$, mediated by neutral pion $(\pi^0)$ exchange. Boson emission in the final state dominates stellar cooling, whereas boson absorption in the initial state dominates superradiance. The remaining diagrams are obtained by attaching the boson to other external neutron legs and by exchanging $p_3$ and $p_4$.}
    \label{fig:FD}
\end{figure}

Within NSs, bosons are efficiently produced through the dominant channel of nucleon – nucleon (NN) bremsstrahlung, $\psi+\psi \to \psi+\psi+\text{boson}$ as well as Cooper pair breaking and formation, see, e.g., Refs.~\cite{Carenza:2019pxu,Shin:2021bvz,Leinson:2014ioa,Hamaguchi:2018oqw,Hamaguchi:2025ztd,Hong:2020bxo,Buschmann:2021juv} and references therein. Figure~\ref{fig:FD} illustrates one of the eight relevant Feynman diagrams for bremsstrahlung, mediated by neutral pion exchange.

The relevant quantity is the emissivity $Q$ due to the NN bremsstrahlung of the boson, expressed as:
\begin{equation}
\begin{aligned}
Q =& \int  \dd \Pi_b \, \omega_b \int \prod^{4}_{i=1}  \dd \Pi_i \, f_1f_2(1-f_3)(1-f_4) \times \delta^4(p_1+p_2-p_3-p_4-p_b) \, (2\pi)^4 \, 
\frac{\sum_s\left|\mathcal{M}_\textrm{SC}\right|^2}{4},
\label{eq:Q}
\end{aligned}
\end{equation}
where $\dd\Pi_{i} \equiv \dd^3 \vec{p}_i/[(2\pi)^3\,2\omega_i]$ denotes the Lorentz-invariant phase space for particle $i$, with $p_i$ and $p_b$ being the neutron and boson momenta, respectively, the Fermi-Dirac distributions $f_i$ describe non-relativistic and degenerate neutrons with Fermi momentum $p_F \approx (3\pi^2 n)^{1/3} \approx 331.5$~MeV, and the factor $1/4$ accounts for identical fermion symmetrization. In Supplemental Material, we provide the full expression for the spin-summed squared amplitude $\sum_s{|\mathcal{M}_\textrm{SC}|^2}$ corresponding to the interactions in Eqs.~(\ref{eq:LaN},\ref{eq:EMDM}).

In the kinetic limit $m_b \ll T \ll p_F \ll m_\psi$, the axion emissivity can be approximated as~\cite{Iwamoto:1984ir,Iwamoto:1992jp,Stoica:2009pion,Harris:2020rus}:
\begin{equation} 
Q_a \approx 10^{-3} \, g_a^2\,m_\psi^2\,T^6\,p_F\,F\left(\frac{m_\pi}{2 p_F}\right)/m_{\pi}^4. 
\end{equation}
where $F(x)$ is a dimensionless function arising from pion exchange, typically of order unity and only weakly dependent on the ratio of the pion mass $m_\pi$ to $2p_F$~\cite{Harris:2020rus}.

 For dark photons with magnetic or electric dipole couplings, the emissivity differs only by replacing $g_a^2$ with $2g_{\rm MDM/EDM}^2$, as shown in Supplemental Material. The factor of $2$ accounts for the two transverse polarization states of the massive vector boson.

The additional boson emissivity must not significantly exceed the standard neutrino emission rate $Q_\nu$~\cite{Friman:1979ecl}, which dominates NS cooling in the early stages. Coupling constraints are obtained by comparing NS cooling simulations that include boson emission with observed luminosities and kinematic ages. This yields a robust bound on the axion-neutron coupling, $g_a \leq 1.3\times10^{-9}$~\cite{Buschmann:2021juv}. Given the similar emissivity structure, these limits rescale directly to dark photon couplings, with $g_{\rm MDM/EDM} \leq 9.2\times10^{-10}$.

\section{Stellar Superradiance from Microphysical Interactions}

In this section, we develop a comprehensive framework for computing the stellar superradiance rate using the same underlying microphysical interactions that govern stellar cooling. We begin by reviewing the general formalism of stellar superradiance arising from boson absorption in rotating stars. We then show how the relevant absorption rate can be computed from microphysical boson-nucleon interactions, emphasizing the long-wavelength suppression induced by multiple neutron scatterings in dense nuclear matter. Finally, we present the resulting superradiance rates and compare them to astrophysical timescales set by NS spin-down measurements.

\subsection{Stellar Superradiance from Absorption}

The absorption of bosons is an essential ingredient in superradiance, as first described by Zeldovich in Ref.~\cite{ZS}. Zeldovich’s mechanism provides a general framework in which a dissipative, rotating object can amplify incident waves, provided the object's angular velocity exceeds the wave's phase velocity. In the rest frame of the stellar medium, dissipation experienced by the bosonic wave arises from the imaginary part of the self-energy and enters the equation of motion as
\be
\left(\Box+m_b^2 + \left(\Gamma^A-\Gamma^E\right) \partial_t \right) \Phi=0,
\label{eq:eomPhi}
\ee
where $\Gamma^A$ and $\Gamma^E$ are the absorption and emission rates obtained from the optical theorem (a detailed derivation using the effective action is given in Supplemental Material). Their difference quantifies the net dissipation. The d’Alembertian $\Box$ incorporates the spacetime metric and thus the gravitational potential well which gives rise to gravitational bound states. We take $\Phi$ to be a complex field whose real part represents either $a$ or $\vec{V}$ in unitary gauge.

Transitioning to an inertial frame in which the object is rotating with angular velocity $\Omega_S$ corresponds to the replacement $\partial_t \rightarrow \partial_t - \Omega_S \partial_\varphi$ in spherical coordinates $(t,r,\theta,\varphi)$. For a bound-state mode with frequency $\omega_b$ and azimuthal number $m$, this introduces the factor $(\omega_b - m\Omega_S)$. When the star rotates fast enough that
\be \omega_b - \Omega_S m < 0,\label{eq:SRC}\ee
the boundstate undergoes exponential growth (superradiance) instead of decay.

For a bosonic field profile bound to a star of mass $M_S$ and radius $R_S$, the system resembles a gravitational atom characterized by the fine-structure constant $\alpha \equiv G_N M_S m_b$~\cite{Detweiler:1980uk,Brito:2015oca}, where $G_N$ is Newton’s constant. We can also define a characteristic (Bohr) radius $r_b = 1/(m_b \alpha)$. In the regime $\alpha \ll 1$, the field profile extends far beyond the stellar radius ($r_b \gg R_S$), allowing a Newtonian approximation for the normalized wavefunctions of the superradiant ground states~\cite{Detweiler:1980uk,Brito:2015oca,Baryakhtar:2017ngi,Chen:2022kzv,Chen:2023vkq}:
\be
\phi = 
\begin{cases}
\displaystyle \frac{1}{8\sqrt{\pi r_b^{3}}}\,\frac{r}{r_b}\,e^{-r/(2r_b)}\,\sin\theta\,e^{-i\omega_b t + i\varphi}, & \text{for } a,\\[8pt]
\displaystyle \frac{1}{\sqrt{2\pi r_b^3}}\,e^{-r/r_b}\,e^{-i\omega_b t}\,(1,\, i,\, 0), & \text{for } \vec{V},
\end{cases}
 \label{eq:phiaV}
\ee
which satisfy the free equation $(\Box + m_b^2)\phi = 0$ with $\omega_b \simeq m_b$ and the normalization  $\int |\phi|^2  \dd ^3  \vec{r} = 1$. Decomposing the field as $\Phi(x) = \Phi_0(t)\,\phi(\vec{r},t)$, with a slowly varying amplitude $\Phi_0$, the amplitude obeys
\be
2\,\omega_b\,\partial_t \Phi_0 \;+\; \langle \Gamma^{\rm net} \rangle\,(\omega_b - m\Omega_S)\,\Phi_0 \;\approx\; 0,
\ee
where 
\be\langle \Gamma^{\rm net} \rangle \equiv \int \phi^*(\Gamma^A - \Gamma^E)\,\phi \, \mathrm{d}^3\vec{r},\ee
denotes the medium-averaged net dissipation inside the star. This yields exponential evolution of the fields total mass ($\propto \Phi_0^2$) with growth rate
\be
\Gamma^{\rm SR} \;=\; \langle \Gamma^{\rm net} \rangle \,\frac{m\Omega_S - \omega_b}{\omega_b}.
\label{eq:growth_rate}
\ee
Thus, for $\Gamma^A>\Gamma^E$ in the medium frame, the mode is damped when $\Omega_S=0$, but grows exponentially once the superradiance condition $\omega_b - m\Omega_S < 0$ is satisfied.

\subsection{Absorption Rate inside Stars}
Boson absorption inside NSs, where the neutron medium is unpolarized and the interactions are those of Eqs.~(\ref{eq:LaN},\,\ref{eq:EMDM}), is dominated by the inverse bremsstrahlung process, in which the boson appears in the initial state, as illustrated in Fig.~\ref{fig:FD}.

The absorption rate is obtained by integrating over the initial and final neutron phase space, weighted by the squared microphysical amplitude, and is given by
\begin{equation}
\begin{aligned}
\Gamma^A =& \frac{1}{2\omega_b} \int \prod^{4}_{i=1} \dd \Pi_i \, f_1f_2(1-f_3)(1-f_4) \times \delta^4(p_1+p_2-p_3-p_4+p_b) \, (2\pi)^4 \, \frac{\sum_{s}\left|\mathcal{M}_{\rm SS}\right|^2}{4}.
\label{eq:GammaA}
\end{aligned}
\end{equation}

The corresponding emission process places the boson with energy $\omega_b$ in the final state, leading to the approximate relation $\Gamma^E \approx e^{-\omega_b/T}\Gamma^A$ for neutrons following a Fermi-Dirac distribution~\cite{Raffelt:1996wa,Noordhuis:2023wid}, and introduces a suppression factor of $\omega_b/T$ to the net dissipation~\cite{Chadha-Day:2022inf,Noordhuis:2023wid}, yielding 
\be \langle\Gamma^{\text{net}}\rangle = \frac{\omega_b}{T} \int  \phi^*\Gamma^A \phi \, \dd^3\vec{r}. \label{eq:Gammanet}
\ee

The amplitude squared for superradiance, $|\mathcal{M}_\textrm{SS}|^2$, involves the same field content as the cooling amplitude. However, the crucial difference lies in the bosonic wavefunctions: cooling utilizes plane-wave states characterized by energies of order $T$, while superradiance employs the bound wavefunctions specified in Eq.~(\ref{eq:phiaV}). In particular, dark photon wavefunctions exhibit fixed macroscopic polarization in superradiance, without polarization summation as required in cooling calculations, as detailed in Supplemental Material.

The non-relativistic nature of the bosonic bound state introduces two sources of suppression in $|\mathcal{M}_{\rm SS}|^2$ relative to the stellar-cooling amplitude $|\mathcal{M}_{\rm SC}|^2$. These arise from (i) collective neutron scattering in the dense medium and (ii) the non-relativistic gradient coupling between the boson and neutrons. For the axion–neutron interaction, the relation takes the form
\be 
|\mathcal{M}^a_{\rm SS}|^2 
\;\approx\;
|\mathcal{M}^a_{\rm SC}|^2\,
F_{\rm sup}\,
\bigg(
\frac{\nabla^{2}}{m_b^{\,2}}
\;+\;
\frac{3 p_F^{2}}{2 m_{\psi}^{2}}
\bigg),
\label{eq:MC}
\ee
where $F_{\rm sup}$ denotes the suppression factor arising from collective neutron scattering, and the Laplacian acts locally on the axion wavefunction $\phi$ in Eq.~(\ref{eq:phiaV}), since the microphysical interaction length is much smaller than both $R_S$ and the Bohr radius $r_b$. Below, we discuss these two suppression mechanisms separately.

\subsubsection{Gravitational-Atom (Bound-State) Suppression}

This suppression originates from the non-relativistic limit of the neutron interactions in Eqs.~(\ref{eq:LaN},\ref{eq:EMDM}), where the neutron spin couples to the spatial gradients of the boson field, such as $\vec{\nabla} a$ for axions and $\vec{\nabla}\!\times\!\vec{V}$ for the MDM coupling of vectors. In contrast, for the EDM interaction the dominant contribution involves $\partial_t \vec{V}$, which does not carry the same gradient suppression~\cite{Chen:2021bdr}. The factor $\sim p_F^{2}/m_\psi^{2}$ in Eq.~(\ref{eq:MC}) reflects the typical neutron relative velocity $p_F/m_\psi \sim 0.3$ in NS matter.

To evaluate the expectation value of the Laplacian operator, we use the local dispersion relation
\be
\left\langle \frac{\nabla^2}{2m_b} \right\rangle  \approx - m_b\,U(r)
= \frac{3G_N M_S m_b}{2R_S}\left(1 - \frac{r^2}{3R_S^2}\right),
\ee
where $U(r)$ is the Newtonian gravitational potential of a uniform-density star. Consequently, the suppression factor in Eq.~(\ref{eq:MC}) is of order $G_N M_S / R_S \sim \mathcal{O}(0.1)$, same order as neutron velocity, instead of the typical $\alpha^2$ associated with the averaged squared velocity of the entire profile, as the NS probes the innermost region of the wavefunction.

\subsubsection{Long-Wavelength Suppression in Dense Nuclear Media}

The suppression factor $F_{\rm sup}$ originates from the fact that, in dense nuclear matter, neutrons undergo many NN scatterings during a single oscillation period of the bosonic bound state. When the boson frequency $\omega_b$ is small compared to the typical NN collision rate, the boson cannot resolve individual scattering events; instead, absorption and emission probe only the collective, rapidly fluctuating neutron medium. This physics is closely analogous to the Landau–Pomeranchuk–Migdal effect in bremsstrahlung processes~\cite{Landau:1953um,Migdal:1956tc}.

A quantitative understanding follows directly from Fig.~\ref{fig:FD} and the standard structure of NN bremsstrahlung in dense matter. Radiation emission or absorption factorizes into an NN scattering kernel and a radiation vertex attached to one of the nucleon legs. In the absence of medium effects, the intermediate nucleon propagator between the NN collision and the radiation vertex is nearly on shell, producing the familiar soft-radiation enhancement $\propto 1/\omega_b$, where $\omega_b$ is the energy transfer carried by the boson. Physically, this corresponds to a formation time $\sim 1/\omega_b$ during which the nucleon resides in a virtual off-shell state.

In a dense medium, however, additional NN scatterings typically occur within this formation time. As shown in classic studies of NN bremsstrahlung~\cite{Friman:1979ecl,Iwamoto:1984ir,Brinkmann:1988vi,Raffelt:1991pw,Raffelt:1996di,Raffelt:1996wa}, these multiple scatterings are incorporated by dressing the intermediate neutron propagator with the imaginary part of its in-medium self-energy, introducing a collision width $\Gamma_{\rm col}$. At the level of the propagator, this amounts to the replacement
\begin{equation}
\frac{1}{\omega_b}\;\longrightarrow\;\frac{1}{\omega_b+i\,\Gamma_{\rm col}}.
\end{equation}

For bosonic absorption or emission, the radiation vertex involves a gradient or time derivative acting on the bosonic field, supplying an explicit factor of $\omega_b$ in the numerator of the amplitude. Combining this with the damped propagator, the squared in-medium amplitude acquires the universal suppression factor
\begin{equation}
F_{\rm sup}
=\left|\frac{\omega_b}{\omega_b+i\Gamma_{\rm col}}\right|^{2}
=\frac{\omega_b^{2}}{\omega_b^{2}+\Gamma_{\rm col}^{2}}.
\end{equation}
In the regime $\omega_b\ll\Gamma_{\rm col}$, many NN collisions occur within a single boson oscillation period, the intermediate neutron state loses phase coherence, leading to strong suppression, $F_{\rm sup}\simeq \omega_b^{2}/\Gamma_{\rm col}^{2}$.

The collision rate for non-degenerate nuclear matter, such as in supernova cores, was estimated in Ref.~\cite{Raffelt:1991pw} using the relation $n\,\sigma v \sim 200~\mathrm{MeV}$, where $\sigma$ is the neutron–neutron cross section and $v\sim p_F/m_\psi$ denotes the typical neutron velocity. However, such estimates do not apply to the cold, highly degenerate neutron medium inside an old NS, where Pauli blocking severely restricts the available scattering phase space.

In the degenerate limit relevant for NSs, the dominant contribution to the in-medium width is the spin–relaxation (collision) rate~\cite{Raffelt:1996di,Raffelt:1996wa},
\begin{equation}\label{eq:GammaDegen}
\Gamma_{\rm col}
= 4\pi\,\alpha_\pi^{2}\,\frac{T^{3}}{p_F^{2}}
\approx 15~\mathrm{meV}\,
\left(\frac{T}{10^{8}\,\mathrm{K}}\right)^{3}
\left(\frac{300~\mathrm{MeV}}{p_F}\right)^{2},
\end{equation}
where the effective one–pion–exchange (OPE) coupling has been taken as $\alpha_\pi = 15$. This rate can also be read off from emissivity calculations for neutrino~\cite{Friman:1979ecl} and axion~\cite{Brinkmann:1988vi,Iwamoto:1984ir} bremsstrahlung emissivities.
The characteristic $T^{3}$ scaling arises from the restricted phase space of quasiparticles within $\mathcal{O}(T)$ of the Fermi surface, leading to a much smaller collision rate than in a non-degenerate medium.

An equivalent parametrization may be obtained from Fermi-liquid theory in terms of the nuclear shear viscosity, yielding quantitatively similar relaxation rates~\cite{Shternin:2013pya,Shternin:2020igy,BaymPethick}. Let $\eta$ denote the shear viscosity of nucleons. Treating the NS matter as a Fermi-Liquid, we have the relation $\eta = \tfrac15 n\,p_F^2\,\tau/m_\psi$ (see, e.g., \cite{Shternin:2013pya,Shternin:2020igy,BaymPethick}). Here, $\tau$ is the relaxation time of the nucleons, which we identify with $\Gamma_{\rm coll}=1/\tau$, giving
\begin{equation}
\Gamma^{\rm Fermi-Liq}_{\rm col}
= \frac{n\,p_F^2}{5\,\eta m_\psi}
\approx 30 ~{\rm meV}\,
\left(\frac{\eta}{10^{20}\ {\rm g\,cm^{-1}s^{-1}}}\right)^{-1}
\left(\frac{n}{n_0}\right)
\left(\frac{p_F}{300~{\rm MeV}}\right)^2, \label{eq:GammaFL}
\end{equation}
where $n_0 \simeq 0.16~{\rm fm^{-3}}$ is nuclear saturation density and we took characteristic values of the shear viscosity $\eta$ from, e.g., \cite{Shternin:2008es}. As expected, both estimates in Eq.~\eqref{eq:GammaDegen} and Eq.~\eqref{eq:GammaFL} are in reasonable agreement.

Although the collision rate in a cold NS is far smaller than the $\mathcal{O}(100\,\mathrm{MeV})$ rates characteristic of supernova matter, it remains orders of magnitude larger than the bosonic frequencies relevant for stellar superradiance, $\omega_b \sim 10^{-14}$–$10^{-11}\,\mathrm{eV}$. Consequently, long-wavelength bosonic absorption is strongly suppressed. By contrast, in stellar cooling the same relaxation rate does not suppress axion bremsstrahlung, since the emitted radiation carries energies $\omega_b \sim T \gg \Gamma_{\rm col}$ in a degenerate NS~\cite{Buschmann:2021juv}.

If nuclear matter enters a superfluid phase, the nucleon mean free path becomes very large, and boson absorption or emission can proceed through Cooper-pair breaking and formation (PBF). At first sight, one might hope that such environments avoid the collision-induced suppression discussed above. However, once superfluidity sets in, different kinematic constraints become dominant. The PBF contribution to boson emissivity in superfluid neutron matter has been computed in Refs.~\cite{Leinson:2014ioa,Hamaguchi:2018oqw,Hamaguchi:2025ztd,Hong:2020bxo,Buschmann:2021juv} and references therein in the context of stellar cooling, where the relevant frequencies satisfy $\omega_b \sim T$.
For absorption through Cooper-pair breaking, the boson must overcome the threshold $\omega_b \ge 2\Delta(T)$, with $\Delta(T)$ the neutron superfluid pairing gap, i.e., the energy required to excite a quasiparticle out of the condensate. Even near the critical temperature $T_c$, one typically has $\Delta(T) \sim \mathcal{O}(10$–$100~\mathrm{keV})$. Thus, while PBF processes are relevant for stellar cooling, they are completely inaccessible to the ultralong-wavelength superradiant modes considered here, for which $\omega_b \sim 10^{-14}$–$10^{-11}~\mathrm{eV}$.

One might also imagine that extremely long-wavelength bosons could couple to global oscillatory modes of the star, such as $r$-modes~\cite{Andersson:2000mf} or other global deformations. However, absorption through such macroscopic degrees of freedom depends sensitively on poorly constrained quantities (e.g. bulk polarization of the medium), and we do not consider these effects further.

\subsection{Superradiance Rates}
We now compute the stellar superradiance rates by evaluating the momentum integral in Eq.~(\ref{eq:GammaA}), followed by the spatial average over the NS volume in Eq.~(\ref{eq:Gammanet}) using the normalized bound-state wavefunctions $\phi$ from Eq.~(\ref{eq:phiaV}). For simplicity, we model the NS as a uniform sphere characterized by temperature $T$ and Fermi momentum $p_F$. Carrying out these steps yields the superradiant growth rates associated with inverse bremsstrahlung, defined in Eq.~(\ref{eq:growth_rate}), for the three types of boson–nucleon couplings:
\begin{align}
   & \Gamma_{a}^{\rm SR}\simeq 5.7\times 10^{-5}g_{a}^2\,C\bigg(\frac{R_S}{r_b}\bigg)^2\left(\frac{G_NM_S}{R_S}+0.7\frac{p_F^2}{m_{\psi}^2}\right) F_{\rm sup},\\
    & \Gamma_{\rm MDM}^{\rm SR}\simeq 8.6\times 10^{-3}g_{\rm MDM}^2\,C\left(\frac{G_NM_S}{R_S}+0.7\frac{p_F^2}{m_{\psi}^2}\right) F_{\rm sup},\\
   & \Gamma_{\rm EDM}^{\rm SR}\simeq 4.0\times 10^{-3}g_{\rm EDM}^2C F_{\rm sup},
   \label{eq:GammaSR3}
\end{align}
where the common factor $C$ is
\begin{equation}
    C\equiv p_F T^2\frac{g_{\pi N}^4m_{\psi}^2}{m_{\pi}^4}\frac{\Omega_S m-\omega_b}{\omega_b}\bigg(\frac{R_S}{r_b}\bigg)^3.
\end{equation}
As emphasized above, each rate is further multiplied by the suppression factor $F_{\rm sup}$ arising from multiple neutron scatterings in the stellar medium.

\subsection{Spin-Down Timescales from Millisecond Pulsars}

Pulsar timing measurements, particularly for millisecond pulsars, provide exquisitely precise determinations of rotation periods and spin-down rates~\cite{1995ASPC72F,2004hpa..book.....L}. Because superradiance leads to exponential extraction of angular momentum, a simple exclusion criterion can be formulated~\cite{Cardoso:2017kgn,Kaplan:2019ako}:
\begin{equation}
    2\,\Gamma^{\rm SR} < \tau_{\rm spindown}^{-1} 
    \equiv \frac{2\,\dot{\Omega}_S}{\Omega_S},
    \label{eq:criterion}
\end{equation}
where $\tau_{\rm spindown}$ denotes the observed spin-down timescale of the pulsar. Note the factor of 2 on the left-hand side of Eq.~\eqref{eq:criterion} arises because $\Gamma^{\rm SR}$ describes growth of the field amplitude, while the spin-down constraint applies to the energy, which scales as amplitude squared. If superradiance were efficient for a given boson mass and coupling, the star would spin down faster than observed.

We examine five of the fastest-known millisecond pulsars, listed in Table~\ref{tab:MSPs}, ordered by decreasing angular velocity and labeled P1–P5. All have rotation frequencies in the kHz range~\cite{atnf_pulsar_group,2005AJ....129.1993M,Stovall:2016unz}. Among them, the second-fastest pulsar, J0952$-$0607 (P2), is especially relevant for superradiance studies. It features a rotation frequency $\Omega_S/(2\pi) \approx 707~\mathrm{Hz}$~\cite{Bassa:2017zpe}, a precisely measured mass $M_S \approx 2.35~M_\odot$~\cite{2022ApJ...934L..17R}, a radius $R_S \approx 14~\mathrm{km}$~\cite{ElHanafy:2023vig}, and an exceptionally long spin-down timescale $\tau_{\rm spindown} \approx 4.7 \times 10^9~\mathrm{yr}$~\cite{Bassa:2017zpe}. For the remaining pulsars, we adopt standard NS parameters $M_S = 1.4~M_\odot$ and $R_S = 12~\mathrm{km}$~\cite{1995ASPC72F}.

\begin{table}[h!]
\centering
\begin{tabular}{c|c|c|c|c}
\hline
 Pulsar& $\Omega_{\rm S}/(2\pi)$[Hz]     & $\tau_{\rm spindown}$[yr]   & $M_S$[$M_{\odot}$]      & $R_S$[km]\\ \hline
P1~\cite{Hessels_2006}    & 716                   &  $>2.5\times10^7$          & $1.4$              & $12$\\ \hline
P2~\cite{Bassa:2017zpe,Ho:2019myl,2022ApJ...934L..17R,ElHanafy:2023vig}      & 707                   &  $4.7\times10^9$            & $2.4$             &$14$\\ \hline
P3~\cite{bkh+82,rsc+21}           &    641                   &  $2.8\times10^8 $      & $1.4$              & $12$ \\ \hline
P4~\cite{hfs+04,EPTA:2023sfo}         &   541                   &  $3.1\times10^9$           & $1.4$              & $12$ \\ \hline
P5~\cite{lbh+15,Stovall:2016unz}      &  380                  &  $5.5\times10^{10}$         & $1.4$              & $12$\\ \hline
\end{tabular}
\caption{Summary of five fast-rotating pulsars selected for superradiance analysis, ordered by angular velocity $\Omega_S$~\cite{atnf_pulsar_group,2005AJ....129.1993M,Stovall:2016unz}, including P1: J1748$-$2446ad~\cite{Hessels_2006}, P2: J0952$-$0607~\cite{Bassa:2017zpe,Ho:2019myl,2022ApJ...934L..17R,ElHanafy:2023vig}, P3: B1937$+$21~\cite{bkh+82,rsc+21}, P4: J1843$-$1113~\cite{hfs+04,EPTA:2023sfo}, and P5: J1938$+$2012~\cite{lbh+15,Stovall:2016unz}. P2 has the largest observed pulsar mass, $M_S \approx 2.35~M_\odot$~\cite{2022ApJ...934L..17R}, and radius $R_S \approx 14$~km~\cite{ElHanafy:2023vig}, while the others are assigned typical values $M_S = 1.4~M_\odot$ and $R_S = 12$~km~\cite{1995ASPC72F}. The spin-down time $\tau_{\rm spindown}$ for P1 is limited by measurement uncertainty, while the others are directly measured.}
\label{tab:MSPs}
\end{table}

Combining the pulsars’ angular velocities with their masses, the superradiant condition in Eq.~(\ref{eq:SRC}) selects a boson mass window near $m_b \sim 10^{-12}\,\mathrm{eV}$ for the dominant superradiant level with azimuthal quantum number $m = 1$, assuming the nonrelativistic limit $\omega_b \approx m_b$. The most rapidly rotating source, P2, can probe values of the gravitational fine-structure parameter up to $\alpha \simeq 0.05$, while the others are sensitive up to $\alpha \simeq 0.03$. Since the pulsars considered have comparable central densities, we take a representative Fermi momentum $p_F \approx 331.5\,\mathrm{MeV}$.

The dominant astrophysical uncertainty arises from the core temperature. While young NSs may be hot, millisecond pulsars are much older objects with expected core temperatures below $10^8~\mathrm{K}$~\cite{Ruderman1983,Ho:2019myl}. In the analysis below, we therefore adopt the range
$T = 10^7-10^8~\mathrm{K}$.

For the most promising source, pulsar P2, we may take the upper bounds on the couplings $g_{a},\, g_{\rm MDM},\, g_{\rm EDM} \sim 10^{-9}$ allowed by existing constraints, and assume a core temperature $T = 10^{8}~\mathrm{K}$. Under these conditions, the maximal superradiant growth for axions occurs at $m_b \simeq 2.6\times 10^{-12}~\mathrm{eV}$, yielding a characteristic timescale $\Gamma_{a}^{-1} \simeq 2.8 \times 10^{9}\,\mathrm{yr}\, F_{\rm sup}^{-1}$. For the MDM and EDM couplings, the optimal boson masses are similar, with corresponding timescales $\Gamma_{\rm EDM}^{-1} \simeq 9.0\times 10^{2}~\mathrm{yr}\, F_{\rm sup}^{-1}$ and $\Gamma_{\rm MDM}^{-1} \simeq 1.3\times 10^{3}~\mathrm{yr}\, F_{\rm sup}^{-1}$.
Thus, in the absence of multiple-scattering effects, the naive free–neutron superradiance rates would be comparable to the observed spin-down timescales of the fastest millisecond pulsars.

However, the true absorption rate is multiplied by the suppression factor
\be
F_{\rm sup} \simeq \frac{m_b^{2}}{\Gamma_{\rm col}^{2}}
    \sim 10^{-16}
    \left(\frac{m_b}{10^{-11}\,\mathrm{eV}}\right)^{2}
    \left(\frac{\mathrm{meV}}{\Gamma_{\rm col}}\right)^{2},\ee
which reduces the effective superradiant growth by many orders of magnitude. As a result, stellar superradiance via inverse bremsstrahlung is astrophysically negligible.

\section{Discussion}

The extreme environments of NSs make them powerful probes of weakly coupled particles beyond the Standard Model, including axions and dark photons. While stellar cooling constrains such couplings through energy loss carried by relativistic bosons, here we have instead examined the extraction of angular momentum from millisecond pulsars via superradiant growth, mediated by the same underlying microphysical interactions. We showed that although the naive superradiance rates obtained from free neutron scattering could, in principle, be competitive with stellar-cooling bounds using measured spin-down rates, additional effects, most importantly multiple nucleon scatterings in the long-wavelength regime, substantially suppress these rates.

Although our analysis has focused on axion and dark-photon couplings to neutrons, similar considerations apply to a broader class of interactions, including quadratic couplings~\cite{Olive:2007aj,Stadnik:2014tta,Stadnik:2015kia,Stadnik:2015xbn,Stadnik:2016zkf,Hees:2018fpg,Hees:2019nhn,Banks:2020gpu,Kim:2022ype,Masia-Roig:2022net,Bouley:2022eer,Banerjee:2022sqg,Beadle:2023flm,Kim:2023pvt,Brax:2023udt,Grossman:2025cov,Gan:2025nlu,Bartnick:2025lbg,Gan:2025icr} and axion–EDM couplings~\cite{Lucente:2022vuo,Springmann:2024ret}. These scenarios would likewise require a careful treatment of dense-matter effects and long-wavelength behavior~\cite{Springmann:2024mjp}.

A more comprehensive framework for the interaction of ultralight fields with dense stellar media, including both normal and superfluid phases, and potentially even global perturbations of the stellar structure, would clearly be valuable. Although the specific channels analyzed in this work yield negligibly small superradiance rates, NSs nonetheless contain enormous reservoirs of rotational energy that dissipate at precisely measured rates. Understanding how new physics might couple to this energy remains an important direction for future study.

If efficient channels for stellar superradiance can be identified, the rich dynamics associated with dense bosonic environments around black holes~\cite{Chen:2019fsq,Yuan:2020xui,Annulli:2020lyc,Chen:2021lvo,East:2022ppo,Chen:2022nbb,Chen:2022kzv,Siemonsen:2022ivj,Chen:2022oad,Chen:2023vkq,Ayzenberg:2023hfw,Spieksma:2023vwl,Aurrekoetxea:2023jwk,Cao:2024wby,Aurrekoetxea:2024cqd,Guo:2024iye,Kim:2024rgf,Tomaselli:2024ojz,Lyu:2025lue,Bai:2025yxm,Kim:2025wwj,Tomaselli:2025zdo,Guo:2025pea,Yu:2025apk,Lyu:2025nsd,Guo:2025ids,Jia:2025vqn,Guo:2025dkx} may likewise manifest in stellar systems, potentially offering a broader range of observational signatures.

\begin{acknowledgments}
We thank Andrea Caputo, Yuxin Liu, and Sam Witte, and especially Georg Raffelt and Damiano Fiorillo for valuable discussions. We also thank Tianyang Gu for help with the code.
The Center of Gravity is a Center of Excellence funded by the Danish National Research Foundation under grant No. DNRF184. V.C. and Y.C. acknowledge support by VILLUM Foundation (grant no. VIL37766) and the DNRF Chair program (grant no. DNRF162) by the Danish National Research Foundation. V.C. is a Villum Investigator and a DNRF Chair.  V.C. acknowledges financial support provided under the European Union's H2020 ERC Advanced Grant “Black holes: gravitational engines of discovery” grant agreement no. Gravitas-101052587.
Views and opinions expressed are however those of the author only and do not necessarily reflect those of the European Union or the European Research Council. Neither the European Union nor the granting authority can be held responsible for them. This project has received funding from the European Union's Horizon 2020 research and innovation programme under the Marie Sklodowska-Curie grant agreement No 101007855 and No 101131233. 
Y.C. is supported by the Rosenfeld foundation in the form of an Exchange Travel Grant and by the COST Action COSMIC WISPers CA21106, supported by COST (European Cooperation in Science and Technology). JIM is grateful for the support of the Science and Technology Facilities Council (STFC) [Grant No. ST/X00077X/1] and funding form a (UKRI) Future Leaders Fellowship [Grant
No. MR/V021974/2].
HS is supported by the Deutsche Forschungsgemeinschaft under
Germany Excellence Strategy — EXC 2121 “Quantum
Universe” — 390833306. 

\end{acknowledgments}


\begin{center}
\textbf{\large Supplemental Material: Stellar Superradiance and Low-Energy Absorption in Dense Nuclear Media}
\end{center}
\appendix
\setcounter{equation}{0}
\setcounter{figure}{0}
\setcounter{table}{0}
\numberwithin{equation}{section}
\counterwithin{figure}{section}
\counterwithin{table}{section}

\makeatletter
\@ifpackageloaded{hyperref}{%
  \renewcommand{\theHsection}{app.\Alph{section}}
  \renewcommand{\theHsubsection}{app.\Alph{section}.\arabic{subsection}}
  \renewcommand{\theHsubsubsection}{app.\Alph{section}.\arabic{subsection}.\arabic{subsubsection}}
  \renewcommand{\theHequation}{app.\Alph{section}.\arabic{equation}}
  \renewcommand{\theHfigure}{app.\Alph{section}.\arabic{figure}}
  \renewcommand{\theHtable}{app.\Alph{section}.\arabic{table}}
}{}
\renewcommand{\bibnumfmt}[1]{[#1]}
\renewcommand{\citenumfont}[1]{#1}
\makeatother

In this Supplemental Material, we present the technical details underlying the dynamics of both stellar cooling (SC) and stellar superradiance (SS). The latter further includes discussions of an effective-action derivation of the boson damping term within the stellar medium.

Given the uncertainties in the neutron star (NS) environment, we adopt benchmark models in which the matter is non-relativistic, degenerate, and non-superfluid. The local properties are characterized by the neutron Fermi momentum $p_F \approx (3\pi^2 n)^{1/3} \approx 331.5$~MeV~\cite{Lattimer:2021emm}, where $n$ is the neutron number density, together with the core temperature $T$. The dominant process in this regime is nucleon--nucleon (NN) bremsstrahlung, $\psi+\psi\rightarrow \psi+\psi+\text{boson}$ and its inverse, where $\psi$ denotes the neutron. The corresponding Feynman diagrams for various neutron-boson couplings defined in Eqs.~(\ref{eq:LaN},\ref{eq:EMDM}) are shown in Fig.~\ref{fig:FD} of the maintext. 

The derivations for both SC and SS follow a common structure:  
\begin{enumerate}
    \item computation of the squared matrix elements for NN bremsstrahlung in the relevant kinematic limits, including angular averaging for each case; and  
    \item integration over the corresponding phase space.
\end{enumerate}

In the non-relativistic and degenerate limit, neutron momenta are approximated by $p_i\equiv(\omega_i,\vec{p_i})\simeq(|\vec{p_i}|^2/(2m_{\psi})+m_{\psi},\vec{p}_i)$, with $|\vec{p}_i| \simeq p_F$ and $i = 1, 2, 3, 4$ labeling the external neutrons in the bremsstrahlung diagrams. The boson momentum is denoted by $p_b \equiv (\omega_b, \vec{p}_b)$, with mass $m_b \ll p_F,\, T$. We model the nuclear interaction using the one-pion exchange (OPE) approximation~\cite{Iwamoto:1984ir,Brinkmann:1988vi,Raffelt:1996wa}, 
noting that significant improvements beyond the simplest OPE calculation have been developed~\cite{Carenza:2019pxu}. The corresponding Lagrangian for the neutral pion-neutron interaction is
\begin{equation}
    \mathcal{L}_{\pi N} = 
    \frac{g_{\pi N}}{m_\pi} 
    (\partial_\mu \pi^0)\, \bar{\psi}\gamma^\mu\gamma^5\psi,
\end{equation}
where $g_{\pi N} \simeq 1.05$ and $m_\pi \simeq 135~\text{MeV}$ is the neutral pion mass. 

For simplicity, we define the neutron momentum transfers as 
$\vec{k} \equiv \vec{p}_2 - \vec{p}_4 \approx \vec{p}_3 - \vec{p}_1$ 
and 
$\vec{l} \equiv \vec{p}_2 - \vec{p}_3 \approx \vec{p}_4 - \vec{p}_1$, 
whose typical magnitudes satisfy 
$|\vec{k}|^2 \sim |\vec{l}|^2 \sim 3 m_{\psi} T$ 
and are much larger than the boson momentum $|\vec{p}_b| \sim T$ 
in the SC regime.
We further introduce the momentum-dependent factors:
\be
\left \{\mathcal{M}_{k}^2  \right \}\equiv\frac{|\vec{k}|^4}{(|\vec{k}|^2+m_{\pi}^2)^2}\,,\\ \left \{\mathcal{M}_{l}^2  \right \}\equiv\frac{|\vec{l}|^4}{(|\vec{l}|^2+m_{\pi}^2)^2}\,,\\ \left \{\mathcal{M}_{kl}^2  \right \}\equiv\frac{|\vec{k}|^2|\vec{l}| ^2}{(|\vec{k}|^2+m_{\pi}^2)(|\vec{l}|^2+m_{\pi}^2)}\,.
\ee

\section{Stellar Cooling}

For the parameter space of interest, the boson mass is much smaller than the stellar temperature, $m_b \ll T$, so the emitted bosons carry typical energy and momentum of order the temperature, $\omega_b \simeq |\vec q_b| \sim T \ll \sqrt{m_{\psi} T}$. We also assume $T \gg \Gamma_{\rm col}$, allowing the neutrons to be treated as effectively free particles in the calculation.

The spin-summed squared matrix elements for the NN bremsstrahlung process with axion or dark-photon emission are
\begin{align}
\sum_{s}|\mathcal{M}^{a}_{\mathrm{SC}}|^2
&= \frac{256}{3}\,
\frac{g_{\pi N}^4\, m_{\psi}^2\, g_a^2}{m_{\pi}^4}
\left(\{\mathcal{M}_{k}^2\} + \{\mathcal{M}_{l}^2\} + \{\mathcal{M}_{kl}^2\}\right), \\[5pt]
\sum_{s}|\mathcal{M}^{\mathrm{EDM/MDM}}_{\mathrm{SC}}|^2
&= \frac{512}{3}\,
\frac{g_{\pi N}^4\, m_{\psi}^2\, g_{\mathrm{EDM/MDM}}^{\,2}}{m_{\pi}^4}
\left(\{\mathcal{M}_{k}^2\} + \{\mathcal{M}_{l}^2\} + \{\mathcal{M}_{kl}^2\}\right),
\end{align}
where the first and second expressions correspond respectively to axion and dark photon (EDM/MDM) couplings.

The relevant observable for stellar cooling is the emissivity, i.e., the energy-loss rate per unit volume, defined as
\begin{align}
Q = \int \dd \Pi_b \, \omega_b 
\!\left[\prod_{i=1}^{4}\! \int \frac{\dd^3\vec{p}_i}{2\omega_i (2\pi)^3}\right]
f_1 f_2 (1 - f_3)(1 - f_4)
(2\pi)^4 \delta^4(p_1 + p_2 - p_3 - p_4 - p_b)
\, \frac{\sum_{s}|\mathcal{M}_{\mathrm{SC}}|^2}{4},
\end{align}
where $f_i = [\exp((\omega_i - \mu_i)/T) + 1]^{-1}$ is the Fermi–Dirac distribution, $\mu_i$ denotes the chemical potential, and the factor $1/4$ accounts for identical initial and final nucleons.  
In the degenerate limit ($T \ll \mu_i$), the dominant contribution arises from neutrons near the Fermi surface with $|\vec{p}_i| \simeq p_F$~\cite{Shapiro:1983du}.  
The nucleon phase-space integrals can be approximated as
\begin{align}
\dd^3\vec{p}_i
= |\vec{p}_i|^2 \dd|\vec{p}_i| \dd \Omega_i
\simeq p_F m_{\psi} \dd \omega_i \dd \Omega_i
= (m_{\psi} p_F T)\, \dd y_i \dd \Omega_i,
\end{align}
where $y_i \equiv (\omega_i - \mu_i)/T$.  
Assuming isotropic boson emission, the angular integration over the boson direction gives a factor of $4\pi$.

The emissivity can be factorized into angular and energy–momentum integrations as
\be\begin{split}
Q_a &=
\frac{p_F^4}{2^{16}\pi^{10}}
\!\left(\frac{256}{3}\frac{g_{\pi N}^4 m_{\psi}^2 g_a^2}{m_{\pi}^4}\right)
A^{\mathrm{SC}} B^{\mathrm{SC}}, \\[3pt]
Q_{\mathrm{EDM/MDM}} &=
\frac{p_F^4}{2^{16}\pi^{10}}
\!\left(\frac{512}{3}\frac{g_{\pi N}^4 m_{\psi}^2 g_{\mathrm{EDM/MDM}}^{\,2}}{m_{\pi}^4}\right)
A^{\mathrm{SC}} B^{\mathrm{SC}},
\end{split}
\ee
corresponding to axion–neutron and dark photon–neutron couplings, respectively.  
Here $A^{\mathrm{SC}}$ and $B^{\mathrm{SC}}$ denote the angular and energy integrals:
\be\begin{split}
A^{\mathrm{SC}} &=
\!\left[\prod_{i=1}^{4}\!\int \dd \Omega_i\right]
\delta^3(\vec{p}_1 + \vec{p}_2 - \vec{p}_3 - \vec{p}_4)
\!\left(\{\mathcal{M}_{k}^2\} + \{\mathcal{M}_{l}^2\} + \{\mathcal{M}_{kl}^2\}\right), \\[5pt]
B^{\mathrm{SC}} &=
T^4 \!\!\int_0^{\infty}\!\!\dd \omega_b \,\omega_b^2
\!\left[\prod_{i=1}^{4}\!\int_{-\mu_i/T}^{\infty}\!\!\dd y_i\right]
\!\delta(\omega_1 + \omega_2 - \omega_3 - \omega_4 - \omega_b)
f_1 f_2 (1 - f_3)(1 - f_4),
\end{split}
\ee
which, in the degenerate limit and assuming chemical equilibrium $\mu_1 + \mu_2 = \mu_3 + \mu_4$, reduces to
\begin{align}
B^{\mathrm{SC}}
&\approx T^6 \!\!\int_0^{\infty}\!\dd y_b\, y_b^2
\!\left[\prod_{i=1}^{4}\!\int_{-\infty}^{\infty}\!\dd y_i\, f_i\right]
\delta\!\left(\sum_i y_i - y_b\right)
= \frac{T^6}{6}\!\int_0^{\infty}\!\dd y_b\, \frac{y_b^3 (y_b^2 + 4\pi^2)}{e^{y_b} - 1}
= \frac{62\pi^6 T^6}{945}.
\end{align}

The angular integral, common to both stellar cooling and superradiance calculations, is given by
\begin{align}
A^{\mathrm{SC}} = 32\pi^3 p_F^{-3} F(\xi),
\end{align}
where $\xi \equiv m_{\pi}/(2p_F)$ and
\begin{align}
F(\xi)
&= 4 - \frac{1}{1+\xi^2}
+ 5\xi \arctan(\xi)
+ \frac{\xi^2}{\sqrt{1+2\xi^2}}
\arctan\!\left(\frac{\xi^2}{\sqrt{1+2\xi^2}}\right)
+ \frac{\pi}{2}\!\left(\frac{\xi}{\sqrt{2\xi^2+1}} - 5\right)\xi.
\label{eq::Fxi}
\end{align}
We provide its detailed derivation in the next section.
This parametrization agrees with full numerical evaluations for temperatures below $10$–$20~\mathrm{MeV}$~\cite{Harris:2020qim}.

\section{Stellar Superradiance}

We now turn to the stellar superradiance (SS) process. The relevant microphysical diagram is the same NN bremsstrahlung diagram that appears in stellar cooling. The key difference lies in the nature of the emitted boson: in SS the ultralight boson occupies a non-relativistic gravitationally bound state. This leads to two important modifications of the calculation.

First, the bound-state wavefunction of the boson must be incorporated when evaluating the rate. Second, the neutron propagator acquires an effective collisional damping width $\Gamma_{\rm col}$, which suppresses the squared matrix element by a factor of order $\omega_b^2 / \Gamma_{\rm col}^2$ due to the much smaller boson energy $\omega_b$. 

For the squared matrix element, we employ two approximations:

\begin{itemize}
\item Light boson limit: $m_b \ll m_{\psi}$ or $\Gamma_{\rm col}$, so the boson mass is negligible compared to the nucleon mass.
\item Non-relativistic regime: both bosons and neutrons have non-relativistic velocities, with $|\vec{q}_b| / m_b$ and $|\vec{p}_i|/m_
\psi\simeq p_F/m_{\psi}$ of order $\mathcal{O}(0.1)$. We treat the magnitudes of these velocities as the small expansion parameters.
We note that in axion dark-matter searches one often assumes the fermion is at rest, so the axion field couples through its spatial gradient alone. However, in SS the relevant object is the relative velocity $\vec{v}_i\equiv \vec{q}_b/m_b-\vec{p}_i/m_{\psi}$.
\end{itemize}

\hspace{1cm}
\begin{center}
\textbf{Axion}    
\end{center}

After performing the two leading-order expansions defined above, we obtain the squared matrix element for axion scattering:
\be\begin{split}
&\sum_{s}\left|\mathcal{M}^{a}_{\rm SS}\right|^2= \Bigg[ 
     \left\{
    \sum_{i=1}^{4}|\vec{v}_i|^2-4(\hat{l}\cdot \vec{v}_1+\hat{l}\cdot \vec{v}_3)(\hat{l}\cdot \vec{v}_2+\hat{l}\cdot \vec{v}_4)-4(\hat{k}\cdot \vec{v}_1+\hat{k}\cdot \vec{v}_3)(\hat{k}\cdot \vec{v}_2+\hat{k}\cdot \vec{v}_4)\right\}\left \{\mathcal{M}_{kl}^2  \right \}\\
    &+\sum_{i=1}^{4}|\vec{v}_i|^2\left(\left \{\mathcal{M}_{k}^2  \right \}+\left \{\mathcal{M}_{l}^2  \right \}\right) +\left\{2(\hat{k}\cdot\vec{v}_1)(\hat{k}\cdot \vec{v}_3)+2(\hat{k}\cdot\vec{v}_2)(\hat{k}\cdot \vec{v}_4)-(\vec{v}_1\cdot\vec{v}_3)-(\vec{v}_2\cdot\vec{v}_4)\right\} \left \{\mathcal{M}_{k}^2  \right \}\\
    &+\left\{2(\hat{l}\cdot\vec{v}_2)(\hat{l}\cdot \vec{v}_3)+2(\hat{l}\cdot\vec{v}_1)(\hat{l}\cdot \vec{v}_4)-(\vec{v}_2\cdot\vec{v}_3)-(\vec{v}_1\cdot\vec{v}_4)\right\} \left \{\mathcal{M}_{l}^2  \right \}\Bigg]\,\frac{g_{\pi N}^4m_{\psi}^2g_{a}^2}{m_{\pi}^4}
    \frac{m_b^2}{\Gamma_{\rm col}^2}.
\end{split}\ee

We then fix the direction of $\vec{q}_b$ and carry out the corresponding angular averages:
\be
\begin{aligned}
   & \braket{\vec{k}\cdot \vec{q}_b}=\braket{\vec{l}\cdot \vec{q}_b}=\braket{\vec{p}_1\cdot \vec{q}_b}=0, \quad \\
   &\braket{(\hat{k}\cdot \hat{q}_b)^2}=\braket{(\hat{l}\cdot \hat{q}_b)^2}=1/3,\\
   & \braket{\vec{l}\cdot \vec{p}_1}= -|\vec{l}|^2/2, \quad \braket{\vec{k}\cdot \vec{p}_1}= -|\vec{k}|^2/2, \\
   &\braket{(\vec{l}\cdot \vec{p}_1)^2}= |\vec{l}|^4/4, \quad \braket{(\vec{k}\cdot \vec{p}_1)^2}= |\vec{k}|^4/4. \label{AA::Axion}
\end{aligned}
\ee
Terms involving $\vec{p}_2$, $\vec{p}_3$, and $\vec{p}_4$ can be rewritten using the definitions of $\vec{k}$ and $\vec{l}$ together with momentum conservation.
After this angular averaging, the squared matrix element for the axion simplifies to
\be
\begin{aligned}
\left\langle\sum_{\text s}\left|\mathcal{M}^{a}_{\rm SS}\right|^2\right\rangle= 
\left[\frac{256}{3}\frac{|\vec{q}_b|^{2}}{m_b^2}+128\frac{p_F^2}{m_{\psi}^2} \right]
\left[\left \{\mathcal{M}_{k}^2  \right \}+\left \{\mathcal{M}_{l}^2  \right \}+\left \{\mathcal{M}_{kl}^2  \right \}\right] \frac{g_{\pi N}^4m_\psi^2\,g_{a}^2}{m_{\pi}^4} \frac{m_b^2}{\Gamma_{\rm col}^2}
.
\label{eq:aampsq}
\end{aligned}
\ee

\hspace{1cm}
\begin{center}
\textbf{MDM Dark Photon}    
\end{center}

For dark photons, a key distinction in the superradiant regime is that the bound state carries a fixed polarization corresponding to a macroscopic spin co-rotating with the star, $\epsilon_{+1}^\mu$.
In computing the squared matrix element, one should therefore use the relation
  \be
    \varepsilon_{+1}^{\mu}(q) \varepsilon_{+1}^{* \nu}(q) \approx \frac{1}{2} \begin{pmatrix}
    0 & 0 & 0 & 0 \\
    0 & 1 & i & 0 \\
    0 & -i & 1 & 0 \\
    0 & 0 & 0 & 0 
\end{pmatrix},
    \ee
where we have taken the non-relativistic limit and neglected the time component.

To simplify the expression, we again perform an angular averages:
\be
 \begin{aligned}
    & \left \langle k_x \,p_{1x}\right \rangle=-\frac{|\vec{k}|^2}{6}, \\ 
    &\left \langle(\vec{k}\cdot\vec{p}_1)\,k_x p_{1x}\right \rangle=\frac{|\vec{k}|^4}{12},\\ 
    &\left \langle (\vec{k}\cdot \vec{p}_1) \,l_x^2\right \rangle=-\frac{|\vec{k}|^2|\vec{l}|^2}{6},\\
    &\left\langle(\hat{k}\cdot\hat{q}_b)^2(l_x^2+l_y^2)\right\rangle=\frac{2}{9}\,|\vec{l}|^2, \\
    &\left\langle(k_x^2+k_y^2)\right\rangle=\frac{2}{3}\,|\vec{k}|^2,\\
    &\braket{k_x l_x}=\braket{k_y l_y}=0
.
\label{AA::DP}
 \end{aligned}
\ee
The final expression follows from the orthogonality condition $\vec{k}\cdot\vec{l} \simeq 0$.
Other required terms can be obtained by the exchange symmetries $x \leftrightarrow y$ and $\vec{k} \leftrightarrow \vec{l}$, and by treating $\vec{p}_2,\,\vec{p}_3,\,\vec{p}_4$ in the same manner as in the axion case. 
We also consider the final volume integral over the boson profile, which renders the distribution of $\vec{q}_b$ isotropic.

The resulting angular-averaged squared matrix element reduces to
\be
  \begin{aligned}
  \left\langle \sum_{s}\big|\mathcal{M}^{\rm MDM}_{\rm SS}\big|^2
  \right\rangle = &
  \;\Bigg\{
 \frac{|\vec{q}_b|^{2}}{m_b^2} \bigg[\frac{4}{3}\{\mathcal{M}_{k}^2\}+\frac{4}{3}\{\mathcal{M}_{l}^2\}+\frac{8}{9}\{\mathcal{M}_{kl}^2\}\bigg] 
 + \frac{2}{3}\frac{p_F^2}{m_{\psi}^2}\big[\{\mathcal{M}_{k}^2\}+\{\mathcal{M}_{l}^2\}\big]  \\
 &+ \frac{7}{6}\left[\frac{|\vec{k}|^2}{m_{\psi}^2}\{\mathcal{M}_{k}^2\} + \frac{|\vec{l}|^2}{m_{\psi}^2}\{\mathcal{M}_{l}^2\}\right] \Bigg\} \frac{64 g_{\pi N}^4 m_\psi^2\, g_{\rm MDM}^2}{m_{\pi}^4}\frac{m_b^2}{\Gamma_{\rm col}^2}.
\end{aligned}
\ee

\hspace{1cm}
\begin{center}
\textbf{EDM Dark Photon}    
\end{center}

For the EDM coupling, the squared matrix element is not suppressed by additional factors of $|\vec{q}_b|^2/m_b^2$ or by $p_f^2/m_{\psi}^2$, in contrast to the MDM case.

The leading-order expansion and angular averaging proceed analogously to the MDM calculation. After performing these steps, the angular-averaged squared matrix element takes the form:
\be
\left\langle
\sum_{s}|\mathcal{M}^{\mrm{EDM}}_{\rm SS}|^2
\right\rangle
=\frac{256}{3} \left(\left \{\mathcal{M}_{k}^2  \right \} +\left \{\mathcal{M}_{l}^2  \right \} +\left \{\mathcal{M}_{kl}^2 \right \} \right) \frac{g_{\pi N}^4m_\psi^2\,g_{\rm EDM}^2}{m_{\pi}^4} \frac{m_b^2}{\Gamma_{\rm col}^2}.
\label{sup edm}
\ee

\subsection{Phase Space Integral}

We now compute the absorption rate $\Gamma^{A}$ for the boson field in the co-rotating frame of the NS. This rate can be written as a phase-space integral over the squared matrix element,
\be
\Gamma^A = \frac{1}{2\omega_b} \left[\prod^{4}_{i=1} \int \frac{\dd^3\vec{p}_i}{2\omega_i(2\pi)^3}  \right]  \, f_1f_2(1-f_3)(1-f_4)\,\delta^4(p_1+p_2-p_3-p_4+p_b) \, (2\pi)^4 \, \frac{\sum_{s}|\mathcal{M}_{\mathrm{SC}}|^2}{4}.
\label{eq:GammaAAppendix}
\ee
As in the stellar-cooling computation, this integral factorizes into angular and energy parts. The resulting rates take the form
\be
\begin{split}
     &\Gamma^{A}_{a}\simeq \frac{p_F^4}{2^{15}\pi^{8}\omega_b}\frac{m_b^2}{\Gamma_{\rm col}^2}\frac{g_{\pi N}^4m_{\psi}^2\,g_{a}^2}{m_{\pi}^4}\left[\frac{256}{3}\frac{|\vec{q}_b|^{2}}{m_b^2}+128\frac{p_F^2}{m_{\psi}^2} \right]A^{\rm SS}B^{\rm SS},\\
      &\Gamma^{A}_{\rm EDM}\simeq \frac{p_F^4}{2^{15}\pi^{8}\omega_b}\frac{m_b^2}{\Gamma_{\rm col}^2}\frac{g_{\pi N}^4m_{\psi}^2\,g_{\rm EDM}^2}{m_{\pi}^4}\frac{256}{3}A^{\rm SS}B^{\rm SS},\\
     &  \Gamma^{A}_{\rm MDM}\simeq \frac{p_F^4}{2^{15}\pi^{8}\omega_b}\frac{m_b^2}{\Gamma_{\rm col}^2}\frac{g_{\pi N}^4m_{\psi}^2\,g_{\rm MDM}^2}{m_{\pi}^4}64\left[\frac{|\vec{q}_b|^{2}}{m_b^2}A^{\rm SS}_1+\frac{p_F^2}{m_{\psi}^2} A^{\rm SS}_2\right]B^{\rm SS}.
\end{split}
\ee
The energy part of the phase-space integral is
\be
\begin{split}
    B^{\rm SS} &=  T^4 \left[\prod_{i=1}^{4}\int_{-\mu_i/T}^{\infty}\dd y_i\right]\,\delta( \omega_1 + \omega_2 - \omega_3 - \omega_4 + \omega_b ) \,f_1f_2(1-f_3)(1-f_4)\\
 &\approx T^3\left[\prod_{i=1}^4\int_{-\infty}^{\infty} \dd y_if_i\right]\delta\bigg(\sum_{i}y_i+y_b\bigg)
 \\
&\approx \frac{T^3}{6} \frac{y_b(y_b^2+4\pi^2)}{1-e^{-y_b}}
\\
&\simeq \frac{2\pi^{2}}{3}T^3,\label{eq:BSS}
\end{split}
\ee
where $y_i \equiv \omega_i/T$. We have used $\mu_1+\mu_2 = \mu_3+\mu_4$ and the limit $\omega_b \ll T$, appropriate for superradiant bosons.
The angular factors are
\be
\begin{aligned}
    A^{\rm SS}_{\rm }=&\left[\prod_{i=1}^{4}\int \dd \Omega_i\right] \,\delta^3 ( \vec{p}_1 + \vec{p}_2 - \vec{p}_3 - \vec{p}_4 ) \left[\left\{\mathcal{M}_{k}^2  \right \} +\left \{\mathcal{M}_{l}^2  \right \} +\left \{\mathcal{M}_{kl}^2 \right \}\right],\\
    A^{\rm SS}_{ 1}=&\left[\prod_{i=1}^{4}\int \dd \Omega_i\right] \,\delta^3 ( \vec{p}_1 + \vec{p}_2 - \vec{p}_3 - \vec{p}_4 ) \left[\frac{4}{3}\{\mathcal{M}_{k}^2\}+\frac{4}{3}\{\mathcal{M}_{l}^2\}+\frac{8}{9}\{\mathcal{M}_{kl}^2\}\right],\\
    A^{\rm SS}_{ 2}=&\left[\prod_{i=1}^{4}\int \dd \Omega_i\right] \,\delta^3 ( \vec{p}_1 + \vec{p}_2 - \vec{p}_3 - \vec{p}_4 ) \left[\frac{2}{3}\bigg(\{\mathcal{M}_{k}^2\}+\{\mathcal{M}_{l}^2\}\bigg)+\frac{7}{6}\bigg(\frac{|\vec{k}|^2}{p_{F}^{2}}\{\mathcal{M}_{k}^2\}+\frac{|\vec{l}|^2}{p_{F}^2}\{\mathcal{M}_{l}^2\}\bigg)\right].
\end{aligned}
\ee

We first evaluate the quantity $A^{\rm SS}$, which is relevant for both the axion and EDM interactions.
A convenient trick is to extend the angular integration measure to full three-momentum integrals by inserting delta functions that fix the momenta on the Fermi surface. This gives
\be
\begin{split}
 A^{\rm SS}
 =& \frac{1}{p_F^8}\left[\prod_{i=1}^{4}\int \dd^3\vec{p}_i\right]  \delta^3 \left( \vec{p}_1 + \vec{p}_2 - \vec{p}_3 - \vec{p}_4 \right)\prod_{i=1}^{4}\delta( |\vec{p}_i|- p_F )
  \left[\left\{\mathcal{M}_{k}^2  \right \} +\left \{\mathcal{M}_{l}^2  \right \} +\left \{\mathcal{M}_{kl}^2 \right \}\right], 
\\
=& \frac{1}{p_F^8}\int \dd^3 \vec{p}_1\, \dd^3 \vec{k} \,\dd^3\vec{l} \,\, \delta \left( |\vec{p}_1| - p_F \right) \delta \left( |\vec{p}_1 + \vec{k} + \vec{l}| - p_F \right)  \delta \left( |\vec{p}_1 + \vec{k}| - p_F \right) \delta \left( |\vec{p}_1 +\vec{l}| - p_F \right)\\
& \times
\left[\left\{\mathcal{M}_{k}^2  \right \} +\left \{\mathcal{M}_{l}^2  \right \} +\left \{\mathcal{M}_{kl}^2 \right \}\right]\\
 \simeq& 16\pi^2p_F^{-4}\int_{0}^{2p_{F}}\dd |\vec{k}| \times 2\int_{0}^{\sqrt{4p_F^2-|\vec{k}|^2}}\dd |\vec{l}|\, \frac{\left\{\mathcal{M}_{k}^2  \right \} +\left \{\mathcal{M}_{l}^2  \right \} +\left \{\mathcal{M}_{kl}^2 \right \}}{\sqrt{4p_{F}^2-|\vec{k}|^2-|\vec{l}|^{2}}}\\ 
 =& 32\pi^{3}p^{-3}_{F} F(\xi),
 \label{eq:A2}
\end{split}
\ee
where $F(\xi)\approx 1.67$ with $\xi=m_{\pi}/(2p_F)$, following the definition in Eq.~(\ref{eq::Fxi}). We have also used the approximate orthogonality $\vec{k}\cdot\vec{l}\simeq 0$.

Similarly,
\be
\begin{split}
    A^{\rm SS}_{\rm 1}\simeq&\,32\pi^3p_F^{-3}F_{\rm MDM}^{1}(\xi)\\
    A^{\rm SS}_{\rm 2}\simeq&\,32\pi^3p_F^{-3}F_{\rm MDM}^{2}(\xi)
\end{split}
\ee
with
\be
\begin{split}
   F_{\rm MDM}^{1}(\xi) &= \frac{2}{9}\left[\pi\bigg(\frac{2\xi}{\sqrt{2\xi^2+1}}-13\bigg)\xi-\frac{6}{\xi^2+1}-\frac{4\xi^2}{\sqrt{1+2\xi^2}}\mrm{arctan}(\frac{\xi^2}{\sqrt{1+2\xi^2}})+26\xi \arctan{\xi}+22\right] \\
  & \approx2.01,\\
   F_{\rm MDM}^{2}(\xi) &=\frac{2}{9(1+\xi^2)}\left[20-105\xi^4-61\xi^2+3(\xi^2+1)(35\xi^2-3)\xi\arctan{\xi}\right]\approx3.36.
\end{split}
\ee

Combining the results above, we obtain
\begin{equation}
\begin{aligned}
  &{\Gamma}^{\rm A}_{a} \simeq \frac{g_{\pi N}^4m_{\psi}^2g_{a}^2}{m_{\pi}^4}\frac{m_b^2}{\Gamma_{\rm col}^2}\left[\frac{256}{3}\frac{|\vec{q}_b|^2}{m_b^2}+128\frac{p_F^2}{m_{\psi}^2}\right]\frac{p_FT^2}{1536\pi^3\omega_b/T}F(\xi), \\ 
  &{\Gamma}^{\rm A}_{\rm MDM} \simeq \frac{g_{\pi N}^4m_{\psi}^2g_{\rm MDM}^2}{m_{\pi}^4}\frac{m_b^2}{\Gamma_{\rm col}^2}\left[\frac{|\vec{q}_b|^2}{m_b^2}F_{\rm MDM}^{1}(\xi)+\frac{p_F^2}{m_{\psi}^2}F_{\rm MDM}^{2}(\xi)\right]\frac{p_FT^2}{24\pi^3\omega_b/T},\\
  & {\Gamma}^{\rm A}_{\rm EDM} \simeq \frac{g_{\pi N}^4m_{\psi}^2g_{\rm EDM}^2}{m_{\pi}^4}\frac{m_b^2}{\Gamma_{\rm col}^2}\frac{1}{18\pi^3}\frac{p_FT^2}{\omega_b/T}F(\xi).
\end{aligned}
\end{equation}

\hspace{1cm}
\subsection{Stellar Superradiance Rates}

Starting from the local absorption rate $\Gamma^{A}$, the net dissipation rate $\langle\Gamma^{\rm net}\rangle$ and the total superradiance growth rate $\Gamma^{\rm SR}$ of the field follows from Eqs.~\eqref{eq:growth_rate} and~\eqref{eq:Gammanet}. In particular,
\be
\Gamma^{\rm SR}=\frac{\omega_b}{T} \frac{\Omega_S m-\omega_b}{\omega_b}
\int\phi^{*}\Gamma_{A}\phi\,\dd^3 \vec{r},
\ee
where $\Gamma_A$ is nonvanishing only within the star.

Carrying out the volume integral over the field profile, the resulting superradiance rates for the three couplings are
\be
\begin{aligned}
   & \Gamma_{a}^{\rm SR}\simeq 5.7\times 10^{-5}g_{a}^2\,C\bigg(\frac{R_S}{r_b}\bigg)^2\left(\frac{G_NM_S}{R_S}+0.7\frac{p_F^2}{m_{\psi}^2}\right) \frac{m_b^2}{\Gamma_{\rm col}^2},\\
    & \Gamma_{\rm MDM}^{\rm SR}\simeq 8.6\times 10^{-3}g_{\rm MDM}^2\,C\left(\frac{G_NM_S}{R_S}+0.7\frac{p_F^2}{m_{\psi}^2}\right) \frac{m_b^2}{\Gamma_{\rm col}^2},\\
   & \Gamma_{\rm EDM}^{\rm SR}\simeq 4.0\times 10^{-3}g_{\rm EDM}^2\,C \frac{m_b^2}{\Gamma_{\rm col}^2},\\
\end{aligned}
\ee
where the common prefactor is
\begin{equation}
    C\equiv p_F T^2\frac{g_{\pi N}^4m_{\psi}^2}{m_{\pi}^4}\frac{\Omega_S m-\omega_b}{\omega_b}\bigg(\frac{R_S}{r_b}\bigg)^3.
\end{equation}

\section{Effective Action Approach to Boson Damping in a Fermion Medium}

In this section, we present the derivation of the damping term that enters the equation of motion of a bosonic field interacting with a fermion medium, using the effective action formalism. We consider a generic bosonic field $\Phi(x)$, which may represent either a scalar or a vector. Its interaction with neutrons is assumed to be of derivative type, written in the compact form
\begin{equation}
\mathcal L_{\text{int}} \;=\; j_A(x)\,\mathcal O^A[\Phi](x),
\end{equation}
where $j_A(x)$ denotes a neutron bilinear (with coupling constants absorbed). In a uniform neutron–star medium, this current should be understood in terms of quasiparticle (dressed) operators. $\mathcal O^A[\Phi](x)$ is a local operator built from $\Phi$ and its derivatives, labeled by the index $A$. We work in the non-relativistic limit for both the bosonic field and the fermion medium. In this regime, the local operator $\mathcal O^A[\Phi]$ may involve spatial or time derivatives, typically appearing as a linear combination of both.

The expectation value of $\Phi$ satisfies the equation of motion
\begin{equation}
\bigl(\Box+m_b^2 \bigr)\,\Phi(x)
\;+\;
\Bigl\langle \frac{\partial \mathcal L_{\text{int}}}{\partial \Phi(x)} \Bigr\rangle
\;=\;0,
\label{eq:EOM-start}
\end{equation}
where the second term involves an ensemble average over the neutron medium. 

The interaction term can be computed by introducing the effective action $\Gamma[\Phi]$ obtained after integrating out the neutron fields, i.e., $\langle {\partial \mathcal L_{\text{int}}}/{\partial \Phi(x)} \rangle = \delta\Gamma/\delta\Phi(x)$. The effective action in the presence of a slowly varying bosonic background can be written in the Schwinger–Keldysh formalism as
\begin{equation}
\Gamma[\Phi] \;=\; \frac{1}{2}
\int \dd^4x\,\dd^4y\;
\mathcal O^A[\Phi](x)\,G^R_{AB}(x-y)\,\mathcal O^B[\Phi](y),
\label{eq:Gamma-nonlocal}
\end{equation}
where
\begin{equation}
G^R_{AB}(x-y)\;=\;-\,i\,\Theta(x^0-y^0)\,\bigl\langle [\,j_A(x),\,j_B(y)\,]\bigr\rangle
\end{equation}
is the retarded correlator of the neutron current $j$, evaluated in the ensemble of the neutron medium. 

Inside a NS, the correlator $G^R_{AB}$ is short-ranged, with a correlation length set by the neutron Fermi momentum. In contrast, the bosonic wavefunction varies only on macroscopic scales, such as the NS radius $R_S$ or the Bohr radius $r_b$. To exploit this separation of scales, we introduce center-of-mass and relative coordinates,
\[
X=\tfrac12(x+y), \qquad \xi=y-x,
\]
and expand $\mathcal O^{A}[\Phi](X\pm \xi/2)$ in a Taylor series around $X$. Retaining only the leading term in this gradient expansion makes the action effectively local in $X$,  
\begin{equation}
\Gamma[\Phi] \;\simeq\; \frac{1}{2}\int \dd^4X\;
\mathcal O^A[\Phi](X)\,\mathcal O^B[\Phi](X)\,G_{AB}(\omega_b),
\label{eq:Gamma-local}
\end{equation}
where the nonlocality has been absorbed into the frequency-space kernel
\begin{equation}
G_{AB}(\omega_b) \;=\; \int \dd^4\xi\; e^{i\omega_b \xi^0}\,G_{AB}^R(\xi).
\end{equation}
Here, the factor $e^{i\omega_b \xi^0}$ arises from the phase difference between $\mathcal O^A[\Phi](X \pm \xi/2)$ and $\mathcal O^A[\Phi](X)$, assuming the bosonic field oscillates coherently at frequency $\omega_b$ with only subdominant spatial variations. In general, $G_{AB}(\omega_b)$ is complex; its real part gives dispersive corrections, while its imaginary part encodes the damping and amplification effects associated with absorption and emission in the neutron medium.

We now examine the imaginary part of $G_{AB}(\omega_b)$. Its microphysical structure becomes explicit upon inserting a complete set of medium eigenstates $\{|i\rangle,|f\rangle\}$ and using the translation property of the dressed current in the NS environment,
$j_A(\xi)=e^{iP\!\cdot\!\xi}\,j_A(0)\,e^{-iP\!\cdot\!\xi}$. In a translation-invariant medium, an approximation valid on length scales shorter than the neutron mean free path, the matrix elements now reduce to a single phase factor, $e^{\,i(E_f-E_i)\xi^0 - i(\vec p_f-\vec p_i)\cdot\vec\xi}$.
In more general backgrounds, the same operator identity still holds exactly, but the resulting matrix elements no longer collapse to a single exponential; instead, they decompose into a sum over multiple momentum modes.

Performing the $\xi$-integration enforces momentum conservation for the fermionic transition. One then obtains
\begin{equation}
\text{Im}\,G_{AB}(\omega_b)
=
-\pi \sum_{i,f}\rho_i\,(2\pi)^3\delta^{3}(\vec p_f-\vec p_i)
\left[
\delta(\omega_b+E_f-E_i)\,M^{AB}_{if}
-
\delta(\omega_b-E_f+E_i)\,M^{AB}_{fi}
\right],
\label{eq:ImG}
\end{equation}
with $M^{AB}_{if}=\langle i|j_A(0)|f\rangle\,\langle f|j_B(0)|i\rangle$ and $\rho_i$ the ensemble weight of $|i\rangle$. The two terms correspond respectively to boson emission ($i\to f+\Phi$) and absorption ($i+\Phi\to f$) in the medium, representing the finite-density optical theorem.

Here the matrix elements $M^{AB}_{if}$ encode the full in-medium response: the currents $j_A$ and $j_B$ are constructed from neutron quasiparticle operators dressed by the surrounding matter, so their matrix elements implicitly incorporate the dispersion and interaction effects characteristic of NS environments.

Substituting Eq.~(\ref{eq:ImG}) into the effective action (\ref{eq:Gamma-local}) and varying then gives the imaginary part of the interaction term in Eq.~(\ref{eq:EOM-start}).

As an example, suppose the operator $\mathcal O^A[\Phi]$ contains a single spatial derivative, i.e., $\mathcal O^i[\Phi] = \partial_i \Phi$. Varying the effective action with respect to $\Phi$ gives
\be \frac{\delta \, \text{Im} \Gamma}{\delta \Phi(x)}
= -\partial_i\!\left[\text{Im} G_{jj}(\omega_b)\,\partial^i \Phi(x)\right] = - \text{Im} G_{jj}(\omega_b)\,\nabla^2 \Phi(x).\label{SMeq:FD}\ee
where the last step assumes $G_{jj}(\omega)$ has much smaller spatial variation than $\Phi$. After integration by parts, the spatial derivative thus generates Laplacian operators acting on $\Phi$ in the effective equation of motion.

By converting the sums in Eq.~(\ref{eq:ImG}) into phase-space integrals, the ensemble weights $\rho_i$ give rise to the thermal occupation factors, such as $f_1 f_2 (1-f_3)(1-f_4)$ in Eq.~(\ref{eq:GammaA}). The factor $\text{Im}\,G_{jj}(\omega_b)\,\nabla^2$ in Eq.~(\ref{SMeq:FD}) can then be identified with the term $(\Gamma^A - \Gamma^E)\,\partial_t$ in Eq.~(\ref{eq:eomPhi}). In this correspondence, the spatial momentum factors that appear in the squared matrix elements are replaced by spatial gradient operators acting on $\Phi$ in position space.

\bibliographystyle{JHEP}
\bibliography{references.bib}

\end{document}